\documentclass[12pt]{iopart}
\usepackage{epsfig}

\begin{document}
\bibliographystyle{unsrt}
 \title[Visibility and non-Markovian effects]{Quantum oscillations in the spin-boson model: Reduced visibility from
non-Markovian effects and initial entanglement}

\author{F.K.\ Wilhelm} 

\address{Institute for Quantum Computing and Department of Physics and 
Astronomy, University of Waterloo, 200 University Avenue W, Waterloo, ON, 
N2L 3G1, Canada}
\ead{fwilhelm@iqc.ca}

%\date{do not distribute this draft of \today}

\begin{abstract}
The loss of coherence of quantum oscillations is of fundamental
interest as well as of practical importance in quantum
computing. In solid-state experiments the oscillations show, next to the
familiar exponential decay on time scales $T_{1/2}$, an overall loss of
amplitude.  We solve the spin-Boson for a large class of initial conditions
without the Markov approximation at the pure dephasing point. It is shown that a loss of visibility occurs
in the form of a fast initial drop for
factorized initial conditions and an overall reduction 
for entangled initial conditions. This loss of amplitude is distict from
$T_2$-decoherence with the difference being most drastic 
for environments with real or pseudo-gaps.  This result is explained by bandwith  effects in
quantum noise as well as in terms of higher-order phase-breaking
processes. For several experiments, such gapped environments are
identified. We conirm that this physics is valid beyond the pure
dephasing point. 

\end{abstract}


\maketitle

\section{Introduction}

Quantum coherence driven by the goal of quantum computation is a
central theme of present-day research in mesoscopic condensed-matter
physics.  In particular in the field of superconducting qubits \cite{Makhlin01,You06,Nato06I,Insight},
spectacular successes have been achieved, such as Rabi oscillations,
charge and flux echo, and a  controlled-not gate
\cite{Nakamura02,Vion02,Chiorescu03,Yamamoto03,Bertet05b,Plantenberg07}. In order to achieve this, the
decoherence due to the ubiquitous  environmental degrees of freedom in
solid-state systems had to be overcome. In fact,
careful modelling of the environment allows to make predictions of the
relaxation and dephasing times, $T_1$ and $T_2$, which are extracted
either from spectroscopic line widths or from fitting exponential
envelopes to Rabi- or Ramsey oscillation data
\cite{PRB03,Ithier05}. Theoretical predictions of $T_1$ are in
reasonable agreement with experiments. $T_2$ is mostly
attributed to $1/f$-noise\cite{Vion02,Ithier05} 
for which self-consistent theories can be formulated \cite{Ithier05,Paladino02}
even though the detailed origins of that noise are 
not quantitatively predictable yet. Technically, the notion of
exponential decay of coherences on scales $T_{1/2}$ is based on a
Markov
approximation of some, potentially implicit, type. 

On the other hand, a 
variety \cite{Vion02,Chiorescu03,Martinis02} of 
controlled experiments show an additional  {\em
loss of visibility}: Next to the
exponential decay given by $T_{1/2}$, the amplitude of coherent
oscillations is reduced even further, i.e., it does not
extrapolate back to the full expected amplitude at $t=0$.
This behavior is not explained
by a simple $T_{1/2}$-picture.  This reduced visibility
is an obstacle for the demonstration of macroscopic quantum
effects in these devices, e.g., of the violation of Bell's inequality
in a coupled qubit system \cite{Steffen06}, and for quantum computing applications.
It has so far mostly been attributed to technological shortcomings
of the detector. This paper introduces a different, additional mechanism
for a reduced
visibility based on non-Markovian effects in the spin boson model. This generic effect is induced i) by higher order
processes involving virtual intermediate states,  which rapidly
entangle system and environment and ii) potential initial entanglement
between system and environment.  We show, that the reduction
originates from the off-resonant high-frequency  parts of the
environmental spectrum which do not contribute to $T_{1/2}$. We
demonstrate that
reduced visibility is compatible with large $T_2$-values in
superohmic and gapped environments and has to 
be considered as an independent quantifier of decoherence. We identify
such environments in recent experiments and estimate
the loss of visibility they induce.

\section{The model}

As a generic model, we consider the
spin-boson Hamiltonian \cite{Leggett87} 
\begin{equation}
H_\theta=\frac{E}{2}\left(\cos\theta\hat{\sigma}_z+\sin\theta\hat{\sigma}_x\right)+\frac12\hat{\sigma}_z\sum_i
\lambda_i (a_i+a^\dagger_i)+\sum_i \omega_i \left(a^\dagger_ia_i+\frac{1}{2}\right).
\label{eq:general_hamiltonian}
\end{equation}
where the $\hat{\sigma}_i$ are Pauli matrices and $a$ and
$a^\dagger$ are Boson annihilation and creation operators. The bath is  characterized by the spectral density $J(\omega)=\sum_i
\left|\lambda_i\right|^2\delta(\omega-\omega_i)$ which is
related to the equilibrium spectral noise power $S(\omega)=J(\omega)\coth(\omega/2T)$. This model is
realized in superconducting qubits, where the bath is the
electromagnetic environment \cite{EPJB03,Nato06II} or by phonons
\cite{Ioffe04}, which also play an important role in quantum dots \cite{Brandes99}. Here and henceforth, we chose
$\hbar=1$ and $k_B=1$. 

The spin-boson model is in general not exactly solvable. It has been
treated with a number of approaches
\cite{Leggett87,Weiss99,DiVincenzo05}.  For quantum computing, the $\lambda_i$ are 
small at $\omega_i\simeq E$ by design and the system-bath coupling 
is usually treated perturbatively. If the system-bath interaction also
defines the longest time in the problem, a Markov approximation
is justified (see, e.g., \cite{Nato06II} for a recent review). This
procedure leads to variants of the well-known Bloch-Redfield master
equation which predicts strictly exponential decay of the spin projections 
contained in the system+bath density matrix $\rho$,
$s_i={\rm Tr} \left[(\sigma_i\otimes 1) \rho\right]$ with time scales
$T_{1/2}$. Thus, conceivably, such an approach cannot describe the loss of visibility. Moreover, many designed environments use
the option to allow for larger $\lambda_i$ at high frequencies\cite{Vion02}, 
$\omega_i\gg E$. 

In order to go beyond Bloch-Redfield, we use two approaches: For
$\theta=0$, the spin-boson model reduces to the exactly solvable
{\em independent boson model} \cite{Mahan00,Shnirman03,Axt05,Muljarov07}.  
For
$\theta\not=0$ we will use perturbation theory in the qubit Hamiltonian to
obtain approximate insights. We will recover similar physics in both cases. 

\section{Pure dephasing point, $\theta=0$}  

We now proceed to the exact solution for a very general initial state in the 
independent boson limit. 
We can perform a Schmid decomposition of the initial density matrix in the
qubit $\otimes$ bath Hilbert space as
\begin{equation}
\rho=\left(\matrix{\rho_{\rm 11}&\rho_{\rm 12}\cr \rho_{\rm 21}&\rho_{\rm 22}}\right).
\end{equation}
As the bath is composed of noninteracting oscillators, the submatrices 
$\rho_{ij}=\prod_k \rho_{ij}^{(k)}$ remain factorized within the bath modes $k$. 
Note that the submatrices do not have to be valid density matrices, only
the complete $\rho$ has to be. 
All matrices with bounded trace can be parameterized with  the characteristic 
function $\chi$ of the Wigner function
$\chi^{(k)}_{ij}(\alpha^{(k)},{\alpha^{(k)}}^\ast)={\rm Tr} \left[\rho^{(k)}_{ij} \hat{D}^{(k)}(-\alpha^{(k)})\right]$ with $\hat{D}^{(k)}(\alpha^{(k)})=\exp(\alpha^{(k)} {a^\dagger}^{(k)} -{\alpha^{\ast}}^{(k)} a_i^{(k)})$ the displacement
operator for mode $k$. Preparing such an initial state requires to use
controls with $\theta(t)\not=0$ at $t<0$.

The fact that for $\theta=0$ the Hamiltonians at different times 
$t>0$ commute allows us to
exactly compute the propagator in interaction representation.
The result can be written as a product of
$D$-operators, $\hat{U}(t)=\prod_k \hat{D}^{(k)}(\hat{\mu}_k(t))$ with
$\hat{\mu}_k
(t)=-\frac{\lambda_k\hat{\sigma}_z}{2\hbar\omega_k}\left(e^{i\omega_k
t}-1\right)$. The application of these to the density matrix is 
straightforward, the essentially induce a coordinate transformation 
on the $\alpha^{(k)}$: The matrix of characteristic functions will be
\begin{eqnarray}
\chi^{(k)}_{ij}(\alpha_{ij}^{(k)})&=&{\rm Tr}_k \left(\rho_{ij}(t=0) D^{(k)}(\mu_i^{(k)})D^{(k)}(-\alpha_{ij}^{(k)}){D^\dagger}^{(k)}(\mu_j^{(k)})\right)\\
&=&e^{i\phi_{ij}^{(k)}}{\rm Tr}_k \left(\rho_{ij}(t=0) D^{(k)}(-\alpha_{ij}^{(k)}+\mu_i^{(k)}-\mu_j^{(k)})\right)\nonumber
\end{eqnarray}
where the phase factor comes from multiplying the displacement
operators and reads $\phi_{ij}^{(k)}={\rm Im} \left[\mu_i{\alpha_{ij}^{(k)}}^\ast+\alpha_{ij}^{(k)}\mu_j^\ast-\mu_i\mu_j^{\ast}\right]$.
We can conclude that 
$\chi^{(k)}_{ij}(\alpha,\alpha^\ast,t)=e^{i\phi_{ij}}\chi^{(k)}_{ij}\left(\alpha_{ij}^{(k)}(t),{\alpha_{ij}^{(k)}}^\ast(t),0\right)$ with $\alpha_{ij}^{(k)}(t)=\alpha_{ij}^{(k)}+\delta\alpha_{ij}^{(k)}(t)$ with $\delta\alpha_{ij}^{(k)}(t)=\mu_i^{(k)}(t)-\mu_j^{(k)}(t)$. This simple property that time evolution is a mere change of
coordinate frame is the main rationale for resorting to $\chi$ as a phase-space
parameterization of the density matrix. The other rationale is that 
$\chi$ makes it in particular easy to compute expectation values of qubit
operators $q$ using the corrolary ${\rm Tr}\left[\rho_{ij}\right] =
{\rm Tr}\left[\rho_{ij}\prod_{k}D^{(k)}_{ij}(0)\right]=\prod_k\chi^{(k)}_{ij}(0,0,t)$ as
\begin{equation}
\langle q\otimes\hat{1}\rangle={\rm Tr} \left[\left(q\otimes\hat{1}\right) \rho\right]
=\sum_{i,j\in\lbrace0,1\rbrace}\langle i|q|j\rangle
\prod_k\chi^{(k)}_{ij}(-\delta\alpha(t),-\delta\alpha^\ast(t),0)
\end{equation}
without any further integration.

From knowing the full density matrix at any time $t>0$ we can calculate any
property of the system we like, including correlation functions. The main 
purpose of this paper is, however, to discuss the quantum system alone in order to establish the connection to master equation approaches. 

Being at the pure dephasing point we can see that there is no dynamics on 
the diagonal elements: $\alpha_{ii}(t)=\alpha_{ii}(0)$ and $\phi_{ii}=0$. 

On the off-diagonal, things get more involved. We focus on studying 
the coherence given in terms of the charateristic Wigner function as
\begin{equation}
s_+=\frac{1}{2}(s_x+is_y)=\prod_k 
\chi_{01}(-\delta\alpha_{01}(t),-\delta\alpha^\ast_{01}(t),0).
\end{equation}

We can work out explicitly
\begin{equation}
\alpha^{(k)}_{01}(t)=\frac{\lambda}{\hbar\omega_i}\left(e^{i\omega_k t}-1\right)
\end{equation}
and it is easy to show that the phase factor drops out. 

We now apply this technique to a physical realistic realization by assuming 
a specific $\chi_{01}(\alpha,\alpha^\ast,0)$. The main restriction 
on this function is that $\rho$ has to be a valid density matrix, i.e. 
Hermitian, normalized, and positive.  
We are restricting ourselves to symmetrically entangling a qubit that is 
initially in an eigenstate of $\sigma_x$ 
with classical states of the bath oscillators, i.e. by choosing for the complete initial density matrix
\begin{equation}
\rho^{(k)}_{ij}=\left(U_D^{(k)}\right)_{ii}\rho^{(k)}_{th}\left({U^{(k)}}^\dagger_D\right)_{jj}
\end{equation}
with $\rho_{th}$ is the thermal density matrix for the bath and 
$U_{D}=D^{(k)}(d^{(k)}\sigma_z)$ displaces each bath mode in opposite directions in 
phase space conditionioned on the two states of the qubit, i.e. it performs a controlled unitary
displacement by some mode-specific amound $d^{(k)}$ of the bath oscillators. Reusing the multiplication property of displacement
operators this means that 
\begin{equation}
\chi_{01}^{(k)}(\alpha,\alpha^\ast,0)=\chi_{\rm th}^{(k)}(\alpha-2d^{(k)},\alpha-2d^{(k)})
\end{equation}
with the phase space function of a thermal state being 
\begin{equation}
\chi_{\rm th}^{(k)}=e^{-2\alpha\alpha^\ast\coth (\omega_k/2T)}
\end{equation}
In this limit, we thus obtain
\begin{equation}
s_{+}=\prod_k e^{-2\coth(\omega_K/2T)|2d_k+\mu_k(t)|^2}.
\end{equation}
taking the continuum limit and substituting
$d_k=(\lambda_k/2\hbar\omega_k)(u_k+iv_k)$ with real
dimensionless coefficients $u_i$ and $v_i$. Taking the continuum
limit we find 
$s_x={\rm Re}\, s_{+}=e^{-K(t)}\cos\epsilon t $ with
\begin{eqnarray}
K(t)&=&-\frac{1}{2}\int_0^\infty \frac{d\omega}{\omega^2}
S(\omega)\left[(u(\omega)+1)^2+v^2(\omega)+1-\right.\\
&&\left.-2\left((1+u(\omega))\cos\omega
t+v(\omega)\sin\omega t\right)\right].\nonumber
\label{eq:gensx}
\end{eqnarray}
Different choices of $u(\omega)$ and $v(\omega)$ can lead to rich
dynamics. We single out prominent quantities: The initial amplitude 
$e^{-K(0)}=\exp\left[-\int
\frac{d\omega}{\omega^2}S(\omega)(u^2(\omega)+v^2(\omega))\right]$,
the time constant $T_2$ of
exponential decay in the long-time  limit,
$1/T_2=\lim_{t\rightarrow\infty}\partial_t K(t)
=S(0)(u(0)+1)$ and the
visibility, the amplitude found when extrapolating the long-time behavior 
back to $t\rightarrow 0$
\begin{equation}
V[u,v]=\lim_{t\rightarrow\infty} e^{\frac{t}{T_2}-K(t)}=e^{-{\cal P}\int
\frac{d\omega}{\omega^2}S(\omega)((1+u(\omega))^2+v^2(\omega))}.
\label{eq:visibility}
\end{equation}
We analyze this
result in important limiting cases.

\subsection{Factorizing initial conditions}

Factorizing initial conditions imply for the density matrix 
 $\rho (t=0)=\rho_{S}\otimes\rho_{B}$, i.e.\ in our notation
$u(\omega)=v(\omega)=0$. This case corresponds to environments which
are part of a detector or any piece of electromagnetic environment
that is switched on {\em during}  the experiment, e.g. the DC-SQUID 
detecting a flux qubit \cite{EPJB03,Burkard05,Bertet05} . It also applies to 
the case of excitons \cite{Huber01,Axt05} as the two-state system 
that are created by ultrafast laser control in the beginning 
of the experiment.
 From the above
result, we can see that this is the only case in which the initial 
amplitude is unity, $K(0)=0$, i.e. when the actual 
oscillations starts at full amplitude.  In this case,
$K(t)$ can be written as \cite{Shnirman03,Sousa06}
\begin{equation}
K(t)=\int_{0}^\infty d\omega  \frac{1-\cos\omega
t}{\omega^2}S(\omega)  =\frac{t^2}{2}\int_{0}^\infty \frac{\rm
sin^2\left(\frac{\omega t}{2}\right)}{(\omega t/2)^2}
 S(\omega).
\label{eq:joft}
\end{equation}
The long-term expansion of this result can be written as
$K(t)=t/T_2+\log V[0,0]+O(1/t)$ with $1/T_2 =S(0)$. Note, that the
rate $1/T_2$is identical to
the Bloch-Redfield result, but additionally the constant term $V[0,0]=\exp\left[-{\cal P}\int_{0}^\infty
\frac{S(\omega)}{\omega^2}\right]$ describes an overall loss of
amplitude. 
Here ${\cal P}$ denotes the Cauchy mean value. 

The dynamics resulting from Eq.\ (\ref{eq:joft}) can be 
interpreted  in terms of environmental quantum noise. This 
is accomplished by looking at the second equality of eq.\
(\ref{eq:joft}) which multiplies $S(\omega)$ with a spectral weight 
of width $1/t$. When the
system has been coupled to the environment for a time $t$, it samples
its spectrum over a bandwidth of $\delta\omega=1/t$ due to 
frequency-time uncertainty. This bandwidth goes to
zero only in the long-time limit.  Quantum-mechanically, the long-time
limit only captures direct energy-conserving processes between
system and environment, whereas at shorter times also higher-order
processes involving energetically  forbidden intermediate states play
a role. The leading process beyond $T_2$ is the excitation of an environmental
mode followed by the relaxation of another, excited state in the environment 
with infinitesimally different energy. These modes have to be distinct
in order to leave a trace in the environment which is a necessary 
condition to
entangle system and environment and hence cause true dephasing.  Thus, it is crucial 
that the environment is in an excited state in the beginning of this cycle. 
In our case, this is guaranteed by the factorized initial conditions: The ground state
of the spin-Boson Hamiltonian \cite{Leggett87} is an appropriate dressed state of the spin. 
When the interaction is switched on, energy is redistributed: Forming 
this dressed state formes lowers the extra energy compared to the
initial state.The extra energy is accomodated in bath
excitations necessary for true dephasing as just explained. 

We will now study the consequences of these results for a number of
important spectra, starting with the ones where $T_2\rightarrow\infty$, 
i.e.\ a constant amplitude at long times. These structured environments 
are an important test-bed for our approach because all the decoherence is 
from non-Markovian effects.
\begin{figure}[ht]
\begin{center}
\includegraphics[width=0.8\columnwidth]{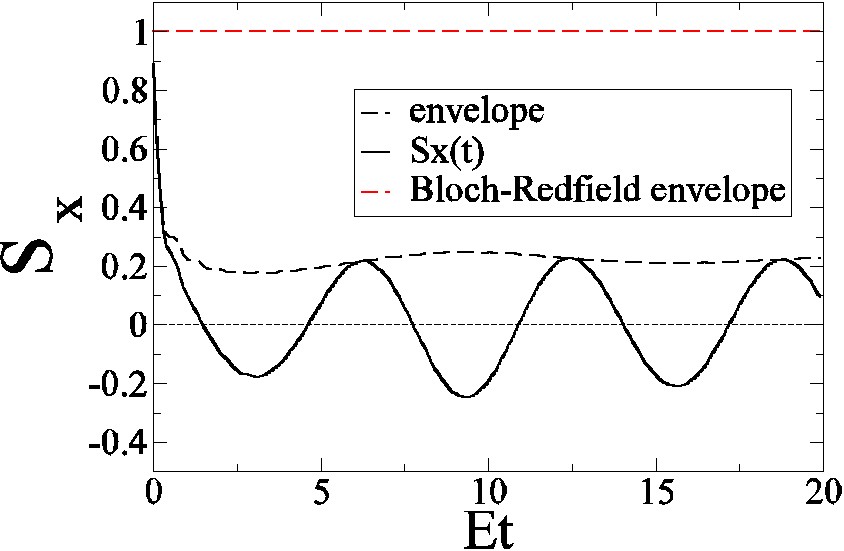}
\end{center}
\caption{Dynamics of $S_x$ and its envelope in the gapped Ohmic model using
$\alpha=0.1$ $\omega_c=10E$, $\omega_{\rm IR}=2E$, and $T=0$. The exact result
is compared to the envelope which would be obtained within Bloch-Redfield theory for
the same model.}
\label{figure_ohmic_slip}
\end{figure}
The gapped Ohmic
model approximately describes the effect of the quasiparticle transport channel
shunting a Josephson junction \cite{Tinkham96}. Its spectral density reads $J_g(\omega)=\alpha_1 \omega
e^{-\omega/\omega_c}\Theta(|\omega|-E_g)$ where $\Theta$ is the
Heaviside unit step function. In the limit of $T\ll E_g\ll\omega_c$,
we find $V[0,0]=\left(\frac{E_g}{\omega_c}\right)^{\alpha_1}
e^{-\alpha_1\gamma}$ where $\gamma\simeq 0.577\dots$ it the
Euler-Mascheroni constant. The amplitude as shown in Fig. \ref{figure_ohmic_slip} drops to this level during a
time $t\simeq 1/E_g$, indicating the time permitted by the energy-time
uncertainty relation over which virtual excitations above the
gap edge may be maintained.

The same physics holds for soft gaps such as in the
standard superohmic case, $J_q(\omega)=\alpha_q \omega^q\omega_c^{1-q}
e^{-\omega/\omega_c}$. This model describes phonon baths as well as
electromagnetic environments blocked off at low frequency by
a serial capacitance \cite{Vion02}. 
Here,  $K(t)$ can be given in closed form \cite{Gorlich88}. For
$q\ge3$ we find a visibility of  $V[0,0]=\exp\left[2\alpha_q\Gamma(q-1)\right]$ for $k_BT\ll \omega_c$. 
For $1<q<3$, the amplitude decays  non-exponentially, also at
long times, given by 
$$e^{-K(t)}\rightarrow\exp\left(2\alpha_q
T\Gamma(q-2) \sin\left(\frac{\pi (q-1)}{2}\right)  \omega_c^{1-q}t^{2-q}\right)$$
and for $q=2$ we find $e^{-K(t)}\rightarrow
e^{-2\alpha_2(1+\kappa)}\left(\frac{t
T}{(1+\kappa)}\right)^{2\alpha_2\kappa}$.  These sub-exponential
decay features cannot be characterized by a time scale
$T_2$ and may resemble reduced visibility if the observation does
not span enough time. 

\begin{figure}[ht]
\begin{center}
\includegraphics[width=0.8\columnwidth]{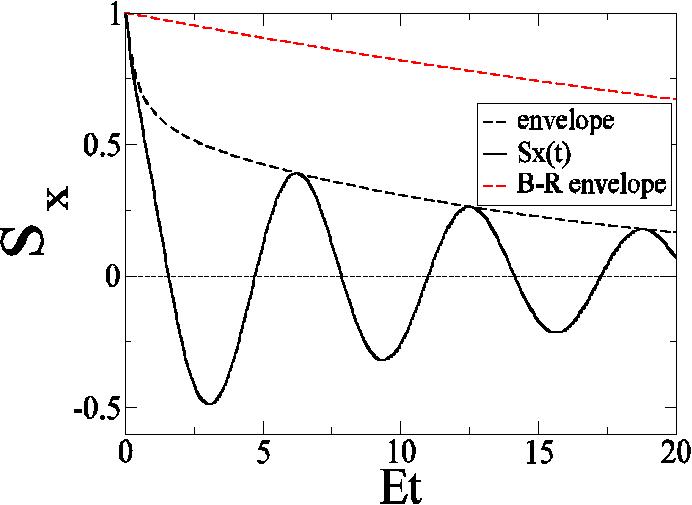}
\end{center}
\caption{Coherent oscillations and their envelope in the Ohmic spin-boson model with
$\omega_c=10\epsilon$, $\alpha=0.1$ and $k_BT=0.5\epsilon$. The envelope
predicted by Bloch-Redfield theory has the correct slope at long times, but misses overall
amplitude.}
\label{figure_super2_slip}
\end{figure}

Loss of visibility is {\em
not} restricted to gapped models. It is usually joined by genuine
exponential
$T_2$ decoherence. This is illustrated in the Ohmic
Spin-Boson model $J_1(\omega)=\alpha_1 \omega e^{-\omega/\omega_c}$,
which describes the quantum version of classical friction and e.g.\
decribes standard resistive environments as well as gapless
electron-hole excitations. At finite $T$,  we identify a finite
$1/T_2=\alpha_1T$. The full shape of the  envelope in the
long-time limit, $tT\gg1$, is however  $e^{-K(t)}\rightarrow
(T/\omega_c)^{\alpha}e^{-t/T_2}$, i.e.\ there is a visibility
prefactor $v[0,0]=(T/\omega_c)^{\alpha}$. At $T=0$,
the long-time limit is never reached as there is no
low-energy scale in the problem, and we find power law decay, 
$e^{-K(t)}=\left(1+\omega_c^2t^2\right)^{-\alpha/2}$. Thus, at $T\rightarrow 0$, 
the very small $v[0,0]$ reflects the fact that for most of the time the 
power-law decay dominates and the long-time expansion involving $T_2$ becomes
valid at extremely long times only. The finite temperature behavior is 
illustrated in Fig. \ref{figure_super2_slip}. Note that this initial
decay followed by exponential decay has been discussed by an entirely
different approach in Ref. \cite{Suarez92}.

%\begin{figure}[ht]
%\begin{center}
%\includegraphics[width=0.9\columnwidth]{./nogaplog.eps}
%\end{center}
%\caption{Loss of visibility in the Ohmic spin-boson model with
%$\omega_c=10\epsilon$ and $\alpha=0.1$. Only envelope
%shown. Extrapolation to short times clearly shows loss of Amplitude.}
%\label{figure_super2_slip}
%\end{figure}

\subsection{Entangled initial condition} 
Factorized initial conditions
are standard assumptions in the theory of open quantum systems,
however, they are not always
realistic. It is already seen in our above result  Eq.\ (\ref{eq:gensx}),
that our prediction of an initial drop 
qualitatively  holds for essentially any
nonequilbrium initial condition. As a complementary case, 
we chose the
$\alpha_i^\pm$ such that in eq.\ (\ref{eq:gensx}) $u_i=-1$,
$v_i=0$. Physically, this state corresponds to a qubit dressed by
environmental oscillators at all frequencies. 
It is the variational (in $u$ and $v$) ground state as long as $E\ll \int d\omega
J(\omega)/\omega$.  It is thus a two-state analog to the 
initial condition discussed in Ref. \cite{Hakim85}.Inspection of eq.\ (\ref{eq:gensx}) shows that these
initial conditions are special in two ways:   i) the envelope $e^{-K(t)}$
is constant in time, $1/T_2=0$ and ii) visibility $V[-1,0]$
has the maximum possible value. It is related to the visibility
of the factorized system by $V[-1,0]=\sqrt{V[0,0]}$.  These
observations can be physically motivated from the minimization of the
total energy, which implies that no further rearrangement of the bath
is necessary for forming the proper dressed states and the dressing is
optimum by accomodating the smallest possible total
energy. Consequently, any other initial condition is not optimal and
contains irreversible parts  of the interaction with the environment,
i.e., genuine $T_2$ decoherence (i.e. finite $T_2$), which we have already identified above as
redistribution of surplus energy.

We appreciate from these results, that the initial conditions play a
decisive role. In fact, the visibility prefactor is a long-time
consequence of short-time physics governed by the initial conditions.
The choice of initial condition depends on the type of physical
environment
and experimental procedure:
A detector is typically switched on during the
experiment and is well described by factorizing initial
conditions. If the switch is non-adiabatic and it happens on a time-scale
$\tau_c$, this can be taken into account by chosing $u(\omega)=0$ for $\omega\tau_c\ll1$ and $u(\omega)=1$ for $\omega\tau_c\gg1$ as an initial state 
in the expression for the visibility. 
The baths  in the material and/or the control environment
are permanently coupled to the qubit.  In a typical experiment, where
on top of the pure dephasing Hamiltonian eq. (\ref{eq:general_hamiltonian})
a large $\Delta\sigma_x$ term is applied at times $t<0$, the entangled
initial condition gives a realistic approximation  for the initial
state, predicting a reduced visibility at all times. Realistic
experiments, involving a non-trivial preparation sequence encoded in
$\theta(t)$ at $t<0$  may be described by variants of these initial
conditions. 
However, the fact that the variational ground state shows
the highest possible visibility outlines the strength of our result:
In many experiments, the detector signal representing $s_x=1$ is not
known {\it a priori} and is obtained by ground state measurements. Following
our results, the visibility obtained in a dynamical experiment is always smaller,
$V[u,v]\le V[-1,0]$.  Thus, our theory does even  apply to these data,
the loss of visibility is not an artefact of the initial condition,
it rather is a generic consequence of the fact that quantum computing
takes place far from thermal equilibrium.
In fact, the short time slip accomodates the extra energy of the system
relative to the dressed ground state. 
Moreover, experiments with good short-time
resolution such as the phase qubit setup show an explicit sign of
an initial drop of the oscillation amplitude \cite{Simmonds04}. 

\section{Beyond pure dephasing}
So far, we have primarily discussed the
exactly solvable pure dephasing point of Hamiltonian eq.\
(\ref{eq:general_hamiltonian}), mainly for being able to easily
circumvent the Born and Markov approximations. Our qualitative results
do however not depend on that assumption. Although a more general solution may
require  more elaborate methods such as path integral expansions \cite{Weiss99}, 
flow equations \cite{Kehrein98,Kehrein05} or real-time RG \cite{Keil01}, 
we can verify this conclusion 
by elementary means for the
gapped environment, when $\omega_{\rm IR}>E$. As shown before, 
this environment does not lead to Markovian decoherence and thus
all decoherence effects are necessarily purely nonmarkovian. 
In this case, it is
legitimate to proceed by perturbation theory in $E/\omega_i\ll 1$.
The eigenstates of the unperturbed, $E=0$, Hamiltonian are two-fold
degenerate and can be written as  $|\psi_\pm,\lbrace
n_k\rbrace\rangle=|\pm\rangle \prod_k D^{(k)}(\pm\lambda_i) |n_i\rangle$ with
eigenenergies  $E(\lbrace n_i \rbrace)=E_0+\sum
(n_i+1/2)\omega_i$. The perturbation Hamiltonian lifts the 
degeneracy and splits the doubletts into sublevels   
\begin{eqnarray}
|g,\lbrace n_i\rbrace\rangle &=&
-\sin\theta_{\rm eff}/2 |+,\lbrace n_i \rbrace\rangle+ \cos\theta_{\rm eff}/2 |-,\lbrace
n_i \rbrace \rangle\\
|e,\lbrace n_i\rbrace\rangle &=& \cos\theta_{\rm eff}/2
|+,\lbrace n_i \rbrace\rangle+ \sin\theta_{\rm eff}/2 |-,\lbrace n_i \rbrace
\rangle.
\end{eqnarray} 
Here, the angle depends on the $n_i$ and is given by
$\tan \theta_{\rm eff}(\lbrace n_i\rbrace)=c(\lbrace n_i\rbrace) \tan\theta$ with $c_i=\left\langle \lbrace n_i\rbrace|D(2\lambda_i)|\lbrace{n_i}\rbrace\right\rangle$. This can be viewed as a down-scaling of the
effective tunnel splitting connected to that specific energies. The shifted energies are $E_{g/e} (\lbrace n_i\rbrace)=\pm E_{\rm eff}(\lbrace n_i\rbrace)/2 +
E(\lbrace n_i\rbrace)$ with $E_{\rm eff}(\lbrace n_i\rbrace)=E\sqrt{\cos^2\theta+\sin^2\theta c(\lbrace n_i\rbrace)^2}$.

We now use these approximate eigenstates for the Gedanken experiment
analogous to the pure dephasing case: We prepare the system in the
equal superposition
$|\psi_0\rangle=\frac{1}{\sqrt{2}}(|g,\lbrace 0\rbrace\rangle+|e,\lbrace 0\rbrace\rangle)$ factorized to
the bath and compute the probability of returning onto the
initial state, $S(t)=|\langle
\psi_0|\psi(t)\rangle|^2$. $|\psi (t)\rangle$ is obtained by expanding the
$|\psi_0\rangle$ in the basis of the approximate eigenstates, 
$|\psi_0\rangle=\sum_i (d_{+}(\lbrace
n_i\rbrace)|e,\lbrace n_i\rbrace\rangle+d_{-}(\lbrace
n_i\rbrace)|g,\lbrace n_i \rbrace\rangle$. The expansion coefficients read
\begin{eqnarray}
d_{-}(\lbrace n_i\rbrace)&=&\sqrt{\frac{\gamma(\lbrace n_i\rbrace)}{2}}
\left(\sin\left(\frac{\theta-\theta_{\rm eff}}{2}\right)+\cos\left(\frac{\theta-\theta_{\rm
          eff}}{2}\right)\right)\\
d_{+}(\lbrace n_i\rbrace)&=&\sqrt{\frac{\gamma(\lbrace n_i\rbrace)}{2}}
\left(-\sin\left(\frac{\theta-\theta_{\rm eff}}{2}\right)+\cos\left(\frac{\theta-\theta_{\rm eff}}{2}\right)\right)
i\end{eqnarray}
where $\gamma=\left\langle\lbrace n_i\rbrace|D(\alpha)|0\right\rangle$.
This expression nicely illustrates our previous point that the initial
state is broken up into entangled states that contain significant 
bath excitations. We now propagate these in time. 
We find without further approximations
\begin{eqnarray}
S(t)&=&\sum_{\lbrace n_i\rbrace}\gamma({\lbrace n_i\rbrace}) e^{-i\sum_i
n_i\omega_i t}\left[\cos (E_{\rm eff}(\lbrace n_i\rbrace) t)+\right.\nonumber\\
&&
\left.\sin(\theta-\theta_{\rm eff})\sin (E_{\rm eff}(\lbrace n_i\rbrace) t)\right]
\end{eqnarray}
We recover the factorizd pure dephasing result in the case of
$\theta=0$. For the other extreme, $\theta=\pi/2$, we recognize 
that the last term is dropping out but the sum cannot be performed
analytically. We observe
because the $\lbrace n_i\rbrace$ dependence of the argument of the cosine provided
additional
decoherence. 
For general $\theta$, the second, phase shifted term that originates
from misalignment of the pure and effective magnetic field
$\theta\not=\theta_{\rm eff}$ on the
Bloch 
sphere provides yet another dephasing channel. A full numerical
analysis of these cases goes far beyond the scope of this paper. 
Qualitatively, we see that the dephasing familiar from the pure
dephasing 
case combines with further dephasing channels, rendering our previous
discussion to remain valid beyond the pure dephasing point. 

Note that $E_{\rm eff}$ at $n_i\equiv0$ leads to the same expression that appears in the adiabatic renormalization
treatment at $\theta=\pi/2$ \cite{Leggett87}. 

Clearly, the results of this section recover the features of the
above discussion and shows, that the assumption of pure dephasing is
not crucial for the physics of the loss of visibility. Any more
quantitative statement should build on more quantitative numerical
methods. A wealth of these methods has been developped in the last
years, including work on the numerical renormalization group
\cite{Anders06}, analytical RG \cite{Kopp07}, and non-Markovian master
equations \cite{Lim06}.

\section{Possible relevance for superconducting qubits}

We have seen that for experiments with long $T_2$, a pronounced loss
of visibility is  governed by gapped baths, unlike the temporal decay, which is dominated
by $1/f$-noise. Such baths can be
identified in experiments. We list a few cases and give their factorized 
visibility. Note that higher values of the visibility indicate the
inadequacy of factorized initial conditions.

  In the flux qubit\cite{Chiorescu03}, the
contribution of the junction quasiparticles can be approximated by a
gapped Ohmic model with  $\omega_{\rm IR}=2\Delta$, $\omega_c=(R_{\rm
N}C_{\rm J})^{-1}$ where $R_{\rm N}$ is the normal-state resistance of
the Josephson junctions and $C_{\rm J}$ the junction capacitance.
$\alpha=R_Q/R_{\rm N}\left(\Phi/\Phi_0-1/2\right)^2$ where  $R_{\rm
Q}=h/4e^2$ is the quantum resistance for Cooper pairs. Using the
numbers of Ref.\ \cite{Chiorescu03} we obtain $V[0,0]=0.8$

In the Quantronium and other samples, the quasiparticles are largely
shunted through the capacitor, $\omega_{\rm IR}>\omega_{\rm
C}$. There, however,  an RC-element is fabricated on chip \cite{Vion02}. The
capacitor is used to decouple the resistor at low frequencies. Indeed,
the environmental spectral density is  super-Ohmic, $J(\omega)\propto
\omega^3$ at $\omega RC\ll 1$. Using the  expression in Ref.\ \cite{Cottet02} with eq.\
\ref{eq:visibility}, we obtain a reduction of visibility from this
term alone of $V[0,0]=0.6$.

Phase qubits exhibit a large number
of spurious resonances. There is no exact mapping onto
an oscillator bath, however, we 
can still estimate a visibility $\log v =-\sum_i \frac{|T_i|^2}{(E-E_i)^2}$ where  $E_i$ is the energy
of the ith resonance and $T_i$ is the coupling matrix element. 
This expression can be 
checked from experimental data when the $T_j$ and $E_j$ have been mapped out. 

Recent experiments in charge qubits report extremely high visibility \cite{Wallraff05}. This
result agrees with our scenario: The off-resonant degrees of freedom which appear in $v$ but not in
$T_{1/2}$ are filtered out by a cavity.

The impact of phonons has been studied in
Ref. \cite{Ioffe04}. Due to the super-Ohmic spectrum they would be a
clear candidate for our scenario, however, the  resulting number even
for factorized initial condition is extremely close to unity.
The topic of visibility has been studied recently for phase qubits
\cite{Meier05,Huang05} and for the Ohmic spin-boson model \cite{DiVincenzo05}. 
The connection between exact solutions and approximate master equations is
discussed in depth in \cite{Doll07}.
Both of these work assume weak coupling to the bath and factorizing initial
conditions. If we expand our
results to lowest order, they cover those papers as special cases. Similar calculations have been
done in Refs.\ \cite{Shnirman03,Axt05} for $\theta=0$ and factorized initial conditions. 

\section{Conclusions}

We have studied the Spin-Boson model beyond the Born and 
Markov approximations. We have shown, that for nonequilibrium initial 
conditions, a long-time observer will measure a reduced visibility of
quantum oscillations as a consequence of short-time physics involving 
higher-order processes. This is well compatible with long $T_2$ for
gapped or superohmic environmental spectra and with a range of experimental
data. 

Discussions with F. Marquardt, S. Kehrein, S. Kohler, D.
Vion, J. Clarke, V. Golovach, J. von Delft, E. Mucciolo, and L. Hollenberg are gratefully acknowledged. Work sponserd by the DFG within SFB 631, NSERC discovery grants, and ARDA through ARO contract P-43385-PH-QC.

\end{document}